\documentclass{lhep}

\journal{LHEP}
\vol{xx}
\jyear{2021}
\pages{xxx}

\received{xx July 2021}
\published{xx July 2021}

\def\be{\begin{equation}}
\def\ee{\end{equation}}
\def\bea{\begin{eqnarray}}
\def\eea{\end{eqnarray}}

\usepackage{graphicx}
\usepackage{dcolumn}
\usepackage{amsfonts}
\usepackage{latexsym}
\usepackage{amssymb}
\usepackage{verbatim}
\usepackage{amsthm}
\usepackage{wrapft}
\usepackage{mathrsfs}
\usepackage{mathptmx}

\newcommand{\NF}{{{\mathbb N}}}

\newcommand{\RF}{{{\mathbb R}}}

\newcommand{\TF}{{{\mathbb T}}}

\newtheorem{theorem}{Theorem}[section]

\newtheorem{proposition}{Proposition}[section]

\newtheorem{conjecture}{Conjecture}[section]
\newtheorem{proposal}{Proposal}[section]

\begin{document}

\title{
  Formal solutions of any-order mass, angular-momentum, dipole
  perturbations on the Schwarzschild background spacetime
}

\author{
  Kouji Nakamura
}
\address{
  Gravitational-Wave Science Project, National Astronomical Observatory of Japan,
  2-21-1, Osawa, Mitaka, Tokyo 181-8588, Japan
}

\begin{abstract}
  Formal solutions of any-order mass, angular-momentum, dipole
  perturbations on the Schwarzschild background spacetime are
  derived in a gauge-invariant manner.
  Once we accept the proposal in [K.~Nakamura, Class. Quantum
  Grav. {\bf 38} (2021), 145010.], we can extend the gauge-invariant
  linear perturbation theory on the Schwarzschild background spacetime including the
  monopole ($l=0$) and dipole ($l=1$) modes to any-order perturbations
  of the same background spacetime through the arguments in
  [K.~Nakamura, Class. Quantum Grav. {\bf 31} (2014), 135013.].
  As a result of this resolution, we reached to a simple derivation of
  the above formal solutions of any order.
\end{abstract}

\maketitle


\section{Introduction}
\label{sec:introduction}


Higher-order perturbation theories are topical subjects in recent
researches on general relativity and they have very wide applications
to cosmology and gravitational-wave physics.
In cosmology, Planck mission revealed the precise map of the
fluctuations of Cosmic Microwave Background
(CMB)~\cite{Planck-Home-Page} and the CMB observation is now regarded
as a precise science.
On the other hand, the direct observation of gravitational waves is
accomplished in 2015~\cite{LIGO-GW150914-2016} and we can expect that
a future direction of gravitational-wave science is also precise
science through the forthcoming data of many gravitational-wave
events.
In addition, some projects of space gravitational-wave antenna are
also progressing~\cite{LISA-homepage,DECIGO-homepage}.
Among them, the Extreme-Mass-Ratio-Inspiral (EMRI), which is a source
of gravitational waves from the motion of a stellar mass object around
a supermassive black hole,  is a promising target of the Laser
Interferometer Space Antenna~\cite{LISA-homepage}.
To describe the gravitational waves from EMRIs, higher-order black
hole perturbation theories are required to support the
gravitational-wave physics as a precise sciences.


In black hole perturbation theories, further sophistication is
possible even in perturbation theories on the Schwarzschild background
spacetime.
There are many studies on the perturbations on the Schwarzschild
background
spacetime~\cite{V.Moncrief-1974a,U.H.Gerlach-U.K.Sengupta-1979a} from
the works by Regge and Wheeler~\cite{T.Regge-J.A.Wheeler-1957} and
Zerilli~\cite{F.Zerilli-1970-PRL}.
In perturbation theories of the Schwarzschild spacetime, we may
decompose the perturbations on this spacetime using the spherical
harmonics $Y_{lm}$ and classify them into odd- and even-modes based on
their parity, because the Schwarzschild spacetime has a spherical
symmetry.
However, monopole ($l=0$) and dipole ($l=1$) modes were separately
treated and their ``{\it gauge-invariant}'' treatments was unknown.


In this situation, in Ref.~\cite{K.Nakamura-2021a}, we proposed a
gauge-invariant treatment of these modes and derived the solutions to
the linearized Einstein equations for these modes.
Since the obtained solutions in Ref.~\cite{K.Nakamura-2021a} is
physically reasonable, we may say that our proposal is also
reasonable.
In addition, owing to our proposal, the formulation of higher-order
gauge-invariant perturbation theory discussed
in~\cite{K.Nakamura-2003,K.Nakamura-2005,K.Nakamura-2011,K.Nakamura-2014}
becomes applicable to any-order perturbations on the Schwarzschild
background spacetime.


In this article, we carry out this application and derive the
formal solutions of mass ($l=0$ even mode), angular momentum ($l=1$
odd mode), and dipole perturbations ($l=1$ even mode) to any-order
perturbations.
We also emphasize that the proposal in Ref.~\cite{K.Nakamura-2021a} is
not only for the perturbations on the Schwarzschild background
spacetime but also a clue  to perturbation theories on a generic
background spacetime such as cosmological perturbation
theories~\cite{K.Nakamura-2006}.


The organization of this paper is as follows:
In Sec.~\ref{sec:review-of-perturbation-theroy}, we briefly review the
framework of the general-relativistic higher-order gauge-invariant
perturbation theory~\cite{K.Nakamura-2003,K.Nakamura-2005,K.Nakamura-2011,K.Nakamura-2014};
In Sec.~\ref{sec:spherical_background_case}, we briefly explain the
strategy for gauge-invariant treatments of $l=0,1$ modes
in Ref.~\cite{K.Nakamura-2021a} and summarize the.
$l=0,1$ mode solutions which was also derived in
Ref.~\cite{K.Nakamura-2021a}.
In Sec.~\ref{sec:higher-order_extension}, we show the extension of the
linear solutions for $l=0,1$ modes to any-order perturbations.
Finally, in Sec.~\ref{sec:Summary}, we provide a brief summary of this
paper.


Throughout this paper, we use the unit $G=c=1$, where $G$ is Newton's
constant of gravitation, and $c$ is the velocity of light.


\section{General-relativistic\\ higher-order gauge-invariant\\ perturbation theory}
\label{sec:review-of-perturbation-theroy}


General relativity is a theory based on general covariance, and that
covariance is the reason that the notion of ``gauge'' has been
introduced into the theory.
In particular, in general relativistic perturbations,
{\it the second-kind gauge} appears in perturbations, as Sachs pointed
out~\cite{R.K.Sachs-1964}.
In general-relativistic perturbation theory, we usually treat the
one-parameter family of spacetimes
$\{({\cal M}_{\lambda},Q_{\lambda})|\lambda\in[0,1]\}$ to discuss
differences between the background spacetime
$({\cal M},Q_{0})$ $=$ $({\cal M}_{\lambda=0},Q_{\lambda=0})$ and the
physical spacetime $({\cal M}_{{\rm ph}},\bar{Q})$ $=$
$({\cal M}_{\lambda=1},Q_{\lambda=1})$.
Here, $\lambda$ is the infinitesimal parameter for perturbations,
${\cal M}_{\lambda}$ is a spacetime manifold for each $\lambda$, and
$Q_{\lambda}$ is the collection of the tensor fields on
${\cal M}_{\lambda}$.
Since each ${\cal M}_{\lambda}$ is a different manifold, we have to
introduce the point identification map ${\cal X}_{\lambda}$ $:$
${\cal M}\rightarrow{\cal M}_{\lambda}$ to compare the tensor field on
different manifolds.
This point-identification is
{\it the gauge choice of the second kind}.
Since we have no guiding principle by which to choose identification
map ${\cal X}_{\lambda}$ due to the general covariance, we may choose
a different point-identification ${\cal Y}_{\lambda}$ from
${\cal X}_{\lambda}$.
This degree of freedom in the gauge choice is {\it the gauge degree of
  freedom of the second kind.}
{\it The gauge-transformation of the second kind} is a change in this
identification map.
We note that this second-kind gauge is a different notion of the
degree of freedom of coordinate choices on a single manifold, which is
called {\it the gauge of the first kind}~\cite{K.Nakamura-2010}.
We have to emphasize that the ``gauge'' which is excluded in our
gauge-invariant perturbation theory is not the gauge of the first kind
but the the gauge of the second kind.
In this paper, we call the gauge of the second kind as {\it gauge} if
there is no possibility of confusions.


Once we introduce the gauge choice ${\cal X}_{k}$ $:$ ${\cal M}$
$\rightarrow$ ${\cal M}_{\lambda}$, we can compare the tensor fields
on different manifolds $\{{\cal M}_{\lambda}\}$, and
{\it perturbations} of a tensor field $Q_{\lambda}$ are represented by
the difference ${\cal X}_{\lambda}^{*}Q_{\lambda} - Q_{0}$, where
${\cal X}_{\lambda}^{*}$ is the pull-back induced by the gauge choice
${\cal X}_{\lambda}$ and $Q_{0}$ is the background value of the
variable $Q_{\lambda}$.
We note that this representation of perturbations completely depends
on the gauge choice ${\cal X}_{\lambda}$.
If we change the gauge choice from ${\cal X}_{\lambda}$ to
${\cal Y}_{\lambda}$, the pulled-back variable of $Q_{\lambda}$ is
then represented by ${\cal Y}_{\lambda}Q_{\lambda}$.
This different representations are related to the gauge-transformation
rules as
\begin{eqnarray}
  \label{eq:gauge-trans-from-calYQ-to-calXQ}
  {\cal Y}_{\lambda}^{*}Q_{\lambda}
  =
  \Phi^{*}_{\lambda} {\cal X}_{\lambda}^{*}Q_{\lambda}
  ,
  \quad
  \Phi_{\lambda} := {\cal X}_{\lambda}^{-1} \circ {\cal Y}_{\lambda}.
\end{eqnarray}
$\Phi_{\lambda}$ is a diffeomorphism on the background spacetime
${\cal M}$.


In the perturbative approach, we treat the perturbation ${\cal
  X}_{\lambda}^{*}Q_{\lambda}$ through the Taylor series with respect
to the infinitesimal parameter $\lambda$ as
\begin{eqnarray}
  \label{eq:calXQ-expand}
  {\cal X}_{\lambda}^{*}Q_{\lambda}
  =:
  \sum_{n=0}^{k} \frac{\lambda^{n}}{n!} {}^{(n)}_{\cal X}\!Q
  +
  O(\lambda^{k+1}),
\end{eqnarray}
where ${}^{(n)}_{\cal X}\!Q$ is the representation associated with the
gauge choice ${\cal X}_{\lambda}$ of the $k$th-order perturbation of
the variable $Q_{\lambda}$ with its background value
${}^{(0)}_{{\cal X}}\!Q=Q_{0}$.
Similarly, we can have the representation of the perturbation of the
variable $Q_{\lambda}$ under the gauge choice ${\cal Y}_{\lambda}$,
which is different from ${\cal X}_{\lambda}$ as mentioned above.
Since these different representations are related to the
gauge-transformation rule (\ref{eq:gauge-trans-from-calYQ-to-calXQ}),
the order-by-order gauge-transformation rule between $n$th-order
perturbations ${}^{(n)}_{\cal X}\!Q$ and ${}^{(n)}_{\cal Y}\!Q$ is
given from the Taylor expansion of the gauge-transformation rule
(\ref{eq:gauge-trans-from-calYQ-to-calXQ}).


Since $\Phi_{\lambda}$ is constructed by the product of
diffeomorphisms, $\Phi_{\lambda}$ is not given by an exponential
map~\cite{K.Nakamura-2003,M.Bruni-S.Matarrese-S.Mollerach-S.Sonego-1997,S.Sonego-M.Bruni-1998},
in general.
For this reason, Sonego and Bruni~\cite{S.Sonego-M.Bruni-1998}
introduced the notion of a {\it knight diffeomorphism} through the
following proposition:
\begin{proposition}
  \label{proposition:K.Nakamura-2014-Proposition2.1}
  Let $\Phi_{\lambda}$ be a one-parameter family of diffeomorphisms,
  and $T$ a tensor field such that $\Phi_{\lambda}^{*}T$ is of class
  $C^{k}$.
  Then, $\Phi_{\lambda}^{*}T$ can be expanded around $\lambda=0$ as
  \begin{eqnarray}
    \label{eq:PhilambdaastT-general-expansion}
    \Phi_{\lambda}^{*}T
    =
    \sum_{n=0}^{k} \lambda^{n}
    \sum_{\{j_{i}\}\in J_{n}} C_{n,\{j_{i}\}}
    {\pounds}_{\xi_{(1)}}^{j_{1}}\cdots{\pounds}_{\xi_{(n)}}^{j_{n}}T
    +
    O(\lambda^{k+1})
    .
  \end{eqnarray}
  Here, $J_{n}:=\left\{ \{j_{i}\} | {}^{\forall}i \in \NF, j_{i}\in
    \NF, s.t. \sum_{i=1}^{\infty} ij_{i}=n\right\}$ defines the set of
  indices over which one has to sum in order to obtain the $n$th-order
  term, $C_{n,\{j_{i}\}}$ $:=$ $\displaystyle \prod_{i=1}^{n}
  \frac{1}{(i!)^{j_{i}}j_{i}!}$, and $O(\lambda^{k+1})$ is a remainder
  with $O(\lambda^{k+1})/\lambda^{k}\rightarrow 0$ in the limit
  $\lambda\rightarrow 0$.
\end{proposition}
The vector fields $\xi_{(1)},...,\xi_{(k)}$ in
Eq.~(\ref{eq:PhilambdaastT-general-expansion}) are called the
generators of $\Phi_{\lambda}$.
The Taylor expansion (\ref{eq:PhilambdaastT-general-expansion}) is a
sufficient representation at least when we concentrate on perturbation
theories~\cite{K.Nakamura-2014,S.Sonego-M.Bruni-1998}.
Actually, this knight diffeomorphism is suitable for our
order-by-order arguments on the gauge issues of general-relativistic
higher-order perturbations.


Through the above notion of the knight diffeomorphism,
Sonego and Bruni also derived the gauge-transformation rules for
$n$th-order perturbations.
As mentioned above, the gauge-transformation rule between the
pulled-back variables ${\cal Y}_{\lambda}^{*}Q_{\lambda}$ and
${\cal X}_{\lambda}^{*}Q_{\lambda}$ is given by
Eq.~(\ref{eq:gauge-trans-from-calYQ-to-calXQ}).
In perturbation theories, we always use the Taylor-expansion of these
variables as in Eq.~(\ref{eq:calXQ-expand}).
To derive the order-by-order gauge-transformation rule for the
$n$th-order perturbation, we have to know the form of the
Taylor-expansion of the pull-back $\Phi_{\lambda}^{*}$ of
diffeomorphism.
Then, we use the general expression
(\ref{eq:PhilambdaastT-general-expansion}) of the Taylor expansion of
diffeomorphisms.
Substituting Eqs.~(\ref{eq:calXQ-expand}) and
(\ref{eq:PhilambdaastT-general-expansion}) into
Eq.~(\ref{eq:gauge-trans-from-calYQ-to-calXQ}), we obtain the
order-by-order expression of the gauge-transformation rules between
the perturbative variables ${}^{(n)}_{{\cal X}}\!Q$ and
${}^{(n)}_{{\cal Y}}\!Q$ as
\begin{eqnarray}
  \label{eq:nth-order-gauge-trans}
  {}^{(n)}_{\;\;{\cal Y}}\!Q - {}^{(n)}_{\;\;{\cal X}}\!Q
  =
  \sum_{l=1}^{n} \frac{n!}{(n-l)!} \sum_{\{j_{i}\}\in J_{l}}
  C_{l,\{J_{i}\}}
  {\pounds}_{\xi_{(1)}}^{j_{1}}
  \cdots
  {\pounds}_{\xi_{(l)}}^{j_{l}}
  {}^{(n-l)}_{\;\;\;\;\;{\cal X}}\!Q
  .
\end{eqnarray}


Inspecting the gauge-transformation rule
(\ref{eq:nth-order-gauge-trans}), we defined gauge-invariant variables
for metric perturbations and for perturbations of an arbitrary tensor
field~\cite{K.Nakamura-2003,K.Nakamura-2005}.
Since the definitions of gauge-invariant variables for perturbations
of an arbitrary tensor field are trivial if we accomplish the
separation of the metric perturbations into their gauge-invariant and
gauge-variant parts, we may concentrate on the metric perturbations,
at first.


We consider the metric $\bar{g}_{ab}$ on the physical spacetime
$({\cal M}_{{\rm ph}},\bar{Q})$ $=$ $({\cal M}_{\lambda=1},Q_{\lambda=1})$,
and we expand the pulled-back metric
${\cal X}_{\lambda}^{*}\bar{g}_{ab}$ to the background spacetime
${\cal M}$ through a gauge choice ${\cal X}_{k}$ as
\begin{eqnarray}
  \label{eq:full-metric-expansion}
  {\cal X}_{\lambda}\bar{g}_{ab}
  =
  \sum_{n=0}^{k} \frac{\lambda^{n}}{n!} {}^{(n)}_{{\cal X}}g_{ab} + O(\lambda^{k+1}),
\end{eqnarray}
where $g_{ab}:={}^{(0)}_{{\cal X}}g_{ab}$ is the metric on the
background spacetime ${\cal M}$.
The expansion (\ref{eq:full-metric-expansion}) of the metric depends
entirely on the gauge choice ${\cal X}_{\lambda}$.
Nevertheless, henceforth, we do not explicitly express the index of
the gauge choice ${\cal X}_{\lambda}$ if there is no possibility of
confusion.
In~\cite{K.Nakamura-2003,K.Nakamura-2005}, we proposed a
procedure to construct gauge-invariant variables for higher-order
perturbations.
Our starting point to construct gauge-invariant variables was the
following conjecture for the linear metric perturbation
$h_{ab}:={}^{(1)}\!g_{ab}$:
\begin{conjecture}
  \label{conjecture:decomposition-conjecture}
  If the gauge-transformation rule for a tensor field $h_{ab}$ is
  given by ${}_{{\cal Y}}\!h_{ab}$ $-$ ${}_{{\cal X}}\!h_{ab}$ $=$
  ${\pounds}_{\xi_{(1)}}g_{ab}$ with the background metric $g_{ab}$,
  there then exist a tensor field ${\cal F}_{ab}$ and a vector
  field $Y^{a}$ such that $h_{ab}$ is decomposed as $h_{ab}$ $=:$
  ${\cal F}_{ab}$ $+$ ${\pounds}_{Y}g_{ab}$, where ${\cal F}_{ab}$ and
  $Y^{a}$ are transformed into ${}_{{\cal Y}}\!{\cal F}_{ab}$ $-$
  ${}_{{\cal X}}\!{\cal F}_{ab}$ $=$ $0$ and ${}_{{\cal Y}}\!Y^{a}$
  $-$ ${}_{{\cal X}}\!Y^{a}$ $=$ $\xi^{a}_{(1)}$ under the gauge
  transformation, respectively.
\end{conjecture}
We call ${\cal F}_{ab}$ and $Y^{a}$ as the
{\it gauge-invariant} and {\it gauge-variant} parts
of $h_{ab}$, respectively.


Based on Conjecture~\ref{conjecture:decomposition-conjecture},
in~\cite{K.Nakamura-2014}, we found that the $n$th-order metric
perturbation ${}^{(n)}_{{\cal X}}g_{ab}$ is decomposed into its
gauge-invariant and gauge-variant parts
as~\footnote{
  Precisely speaking, to reach to the decomposition formula
  (\ref{eq:nth-order-original-ngab-decomp}), we have to confirm
  Conjecture 4.1 in Ref.~\cite{K.Nakamura-2014} in addition to
  Conjecture~\ref{conjecture:decomposition-conjecture}.
}
\begin{eqnarray}
  {}^{(n)}\!g_{ab}
  \!\!\!\!\!\!\!\!
  &
    =\!\!\!\!\!\!\!\!
  &
    {}^{(n)}\!{\cal F}_{ab}
    \nonumber\\
  &&
     -
     \sum_{l=1}^{n}
     \frac{n!}{(n-l)!}
     \sum_{\{j_{i}\}\in J_{l}}
     C_{l,\{j_{i}\}}
     {\pounds}_{-{}^{(1)}\!Y}^{j_{1}}\cdots{\pounds}_{-{}^{(l)}\!Y}^{j_{l}}
     {}^{(n-l)}\!g_{ab}
     .
     \label{eq:nth-order-original-ngab-decomp}
\end{eqnarray}
Furthermore, through the gauge-variant variables ${}^{(i)}Y^{a}$
($i=1,...,n$), we also found the definition of the gauge-invariant
variable ${}^{(n)}\!{\cal Q}$ for the $n$th-order perturbation
${}^{(n)}\!Q$ of an arbitrary tensor field $Q$.
This definition of the gauge-invariant variable ${}^{(n)}\!{\cal Q}$
implies that the $n$th-order perturbation ${}^{(n)}\!Q$ of any tensor
field $Q$ is always decomposed into its gauge-invariant part
${}^{(n)}\!{\cal Q}$ and gauge-variant part as
\begin{eqnarray}
  {}^{(n)}\!Q
  =
  {}^{(n)}\!{\cal Q}
  -
  \sum_{l=1}^{n}
  \frac{n!}{(n-l)!}
  \sum_{\{j_{i}\}\in J_{l}}
  C_{l,\{j_{i}\}}
  {\pounds}_{-{}^{(1)}\!Y}^{j_{1}}\cdots{\pounds}_{-{}^{(l)}\!Y}^{j_{l}}
  {}^{(n-l)}\!Q
  .
  \label{eq:nth-order-original-nQ-decomp}
\end{eqnarray}


As an example, the perturbative expansion of the Einstein tensor and
the energy-momentum tensor, which are pulled back through the gauge
choice ${\cal X}_{\lambda}$, are given by
\begin{eqnarray}
  \label{eq:barGab-expansion}
  {\cal X}_{\lambda}^{*}\bar{G}_{a}^{\;\;b}
  &=&
      \sum_{n=0}^{k} \frac{\lambda^{n}}{n!}
      {}^{(n)}_{{\cal X}}\!G_{a}^{\;\;b}
      +
      O(\lambda^{k+1})
      ,
  \\
  \label{eq:arTab-expansion}
  {\cal X}_{\lambda}^{*}\bar{T}_{a}^{\;\;b}
  &=&
      \sum_{n=0}^{k} \frac{\lambda^{n}}{n!}
      {}^{(n)}_{{\cal X}}\!T_{a}^{\;\;b}
      +
      O(\lambda^{k+1})
      .
\end{eqnarray}
Then, the $n$th-order perturbation
${}^{(n)}_{{\cal X}}G_{a}^{\;\;b}$ of the Einstein tensor and the
$n$th-order perturbation ${}^{(n)}_{{\cal X}}T_{a}^{\;\;b}$ of the
energy-momentum tensor are also decomposed as
\begin{eqnarray}
  {}^{(n)}\!G_{a}^{\;\;b}
  \!\!\!\!\!\!\!\!
  &
    =\!\!\!\!\!\!\!\!
  &
    {}^{(n)}\!{\cal G}_{a}^{\;\;b}
    \nonumber\\
  &&
     -
     \sum_{l=1}^{n}
     \frac{n!}{(n-l)!}
     \sum_{\{j_{i}\}\in J_{l}}
     C_{l,\{j_{i}\}}
     {\pounds}_{-{}^{(1)}\!Y}^{j_{1}}\cdots{\pounds}_{-{}^{(l)}\!Y}^{j_{l}}
     {}^{(n-l)}\!G_{a}^{\;\;b}
     ,
     \label{eq:nth-order-original-nGab-decomp}
  \\
  {}^{(n)}\!T_{a}^{\;\;b}
  \!\!\!\!\!\!\!\!
  &
    =\!\!\!\!\!\!\!\!
  &
    {}^{(n)}\!{\cal T}_{a}^{\;\;b}
    \nonumber\\
  &&
     -
     \sum_{l=1}^{n}
     \frac{n!}{(n-l)!}
     \sum_{\{j_{i}\}\in J_{l}}
     C_{l,\{j_{i}\}}
     {\pounds}_{-{}^{(1)}\!Y}^{j_{1}}\cdots{\pounds}_{-{}^{(l)}\!Y}^{j_{l}}
     {}^{(n-l)}\!T_{a}^{\;\;b}
     .
     \label{eq:nth-order-original-nTab-decomp}
\end{eqnarray}
Through the lower-order Einstein equation
${}^{(k)}_{{\cal X}}\!G_{a}^{\;\;b}=8\pi{}^{(k)}_{{\cal X}}\!T_{a}^{\;\;b}$
with $k\leq n-1$, the $n$th-order Einstein equation
${}^{(n)}_{{\cal X}}\!G_{a}^{\;\;b}=8\pi{}^{(n)}_{{\cal X}}\!T_{a}^{\;\;b}$
is automatically given in the gauge-invariant form
\begin{eqnarray}
  {}^{(n)}\!{\cal G}_{a}^{\;\;b} = 8\pi {}^{(n)}\!{\cal T}_{a}^{\;\;b}.
  \label{eq:nth-order-Einstein-eq}
\end{eqnarray}
Here, we note that the $n$th-order perturbation of the Einstein tensor
is given in the form
\begin{eqnarray}
  \label{eq:nth-einstein-tensor-separate}
  {}^{(n)}\!{\cal G}_{a}^{\;\;b}
  =
  {}^{(1)}\!{\cal G}_{a}^{\;\;b}\left[{}^{(n)}\!{\cal F}\right]
  +
  {}^{({\rm NL})}\!{\cal G}_{a}^{\;\;b}\left[\left\{\left.{}^{(i)}\!{\cal F}\right|i<n\right\}\right]
  ,
\end{eqnarray}
where ${}^{(1)}\!{\cal G}_{a}^{\;\;b}$ is the gauge-invariant part of
the linear-order perturbation of the Einstein tensor.
Explicitly, ${}^{(1)}\!{\cal G}_{a}^{\;\;b}\left[A\right]$ for an
arbitrary tensor field $A_{ab}$ of the second rank is given
by~\cite{K.Nakamura-2005,K.Nakamura-2010}
\begin{eqnarray}
  \label{eq:linear-Einstein-AIA2010-2}
  \!\!\!\!\!\!\!\!\!\!\!\!\!\!\!\!
  &&
     {}^{(1)}{\cal G}_{a}^{\;\;b}\left[A\right]
     :=
     {}^{(1)}\Sigma_{a}^{\;\;b}\left[A\right]
     - \frac{1}{2} \delta_{a}^{\;\;b} {}^{(1)}\Sigma_{c}^{\;\;c}\left[A\right]
     ,
  \\
  \label{eq:(1)Sigma-def-linear}
  \!\!\!\!\!\!\!\!\!\!\!\!\!\!\!\!
  &&
     {}^{(1)}\Sigma_{a}^{\;\;b}\left[A\right]
     :=
     - 2 \nabla_{[a}^{}H_{d]}^{\;\;\;bd}\left[A\right]
     - A^{cb} R_{ac}
     ,
  \\
  \label{eq:Habc-def-linear}
  \!\!\!\!\!\!\!\!\!\!\!\!\!\!\!\!
  &&
     H_{ba}^{\;\;\;\;c}\left[A\right]
     :=
     \nabla_{(a}A_{b)}^{\;\;\;\;c} - \frac{1}{2} \nabla^{c}A_{ab}
     .
\end{eqnarray}
As derived in~\cite{K.Nakamura-2005}, when the background Einstein
tensor vanishes, we obtain the identity
\begin{eqnarray}
  \label{eq:linear-perturbation-of-div-Gab-vacuum}
  \nabla_{a}{}^{(1)}\!{\cal G}_{b}^{\;\;a}\left[A\right]
  =
  0
\end{eqnarray}
for an arbitrary tensor field $A_{ab}$ of the second rank.


Thus, we emphasize that
Conjecture~\ref{conjecture:decomposition-conjecture} was the important
premise of the above framework of the higher-order perturbation theory.


\section{Linear perturbations on\\ the Schwarzschild\\ background spacetime}
\label{sec:spherical_background_case}


We use the 2+2 formulation~\cite{U.H.Gerlach-U.K.Sengupta-1979a} of
the perturbations on spherically symmetric background spacetimes.
The topological space of spherically symmetric spacetimes is the
direct product ${\cal M}={\cal M}_{1}\times S^{2}$, and the metric on
this spacetime is
\begin{eqnarray}
  &&
     g_{ab}
     =
     y_{ab} + r^{2}\gamma_{ab}
     ,
     \label{eq:background-metric-2+2}
     \\
  &&
     y_{ab} = y_{AB} (dx^{A})_{a}(dx^{B})_{b}
     ,
     \;\;\;
     \gamma_{ab} = \gamma_{pq} (dx^{p})_{a} (dx^{q})_{b}
     ,
     \label{eq:background-metric-2+2-separate}
\end{eqnarray}
where $x^{A} = (t,r)$ and $x^{p}=(\theta,\phi)$.
In addition, $\gamma_{pq}$ is a metric of the unit sphere.
In the Schwarzschild spacetime, the metric
(\ref{eq:background-metric-2+2}) is given by
\begin{eqnarray}
  \label{eq:background-metric-2+2-y-comp-Schwarzschild}
  &&
     y_{ab}
     =
     - f (dt)_{a}(dt)_{b}
     +
     f^{-1} (dr)_{a}(dr)_{b}
     ,
  \\
  \label{eq:background-metric-2+2-f-Schwarzschild}
  &&
     f = 1 - \frac{2M}{r}
     ,
  \\
  \label{eq:background-metric-2+2-gamma-comp-Schwarzschild}
  &&
     \gamma_{ab}
     =
     (d\theta)_{a}(d\theta)_{b}
     +
     \sin^{2}\theta(d\phi)_{a}(d\phi)_{b}
     .
\end{eqnarray}


On this background spacetime $({\cal M},g_{ab})$, we consider the
components of the metric perturbation as
\begin{eqnarray}
  &&
     h_{ab}
     =
     h_{AB} (dx^{A})_{a}(dx^{B})_{b}
     +
     2 h_{Ap} (dx^{A})_{(a}(dx^{p})_{b)}
     \nonumber\\
  && \quad\quad\quad
     +
     h_{pq} (dx^{p})_{a}(dx^{q})_{b}
     .
\end{eqnarray}
In Ref.~\cite{K.Nakamura-2021a}, we proposed the decomposition of
these components as
\begin{eqnarray}
  \label{eq:hAB-fourier}
  &&
     h_{AB}
     =
     \sum_{l,m} \tilde{h}_{AB} S_{\delta}
     ,
  \\
  &&
     \label{eq:hAp-fourier}
     h_{Ap}
     =
     r \sum_{l,m} \left[
     \tilde{h}_{(e1)A} \hat{D}_{p}S_{\delta}
     +
     \tilde{h}_{(o1)A} \epsilon_{pq} \hat{D}^{q}S_{\delta}
     \right]
     ,
  \\
  &&
     \label{eq:hpq-fourier}
     h_{pq}
     =
     r^{2} \sum_{l,m} \left[
     \frac{1}{2} \gamma_{pq} \tilde{h}_{(e0)} S_{\delta}
     +
     \tilde{h}_{(e2)} \left(
     \hat{D}_{p}\hat{D}_{q} - \frac{1}{2} \gamma_{pq} \hat{\Delta}
     \right) S_{\delta}
     \right.
     \nonumber\\
  && \quad\quad\quad\quad\quad\quad
     \left.
     +
     2 \tilde{h}_{(o2)} \epsilon_{r(p} \hat{D}_{q)}\hat{D}^{r} S_{\delta}
     \right]
     ,
\end{eqnarray}
where $\hat{D}_{p}$ is the covariant derivative associated with
the metric $\gamma_{pq}$ on $S^{2}$,
$\hat{D}^{p}:=\gamma^{pq}\hat{D}_{q}$, and
$\epsilon_{pq}=\epsilon_{[pq]}$ is the totally antisymmetric
tensor on $S^{2}$.


Note that the decomposition
(\ref{eq:hAB-fourier})--(\ref{eq:hpq-fourier}) implicitly state that
the Green functions of the
derivative operators $\hat{\Delta}:=\hat{D}^{r}\hat{D}_{r}$ and
$\hat{\Delta}+2:=\hat{D}^{r}\hat{D}_{r}+2$ should exist if the
one-to-one correspondence between $\{h_{Ap},$ $h_{pq}\}$ and
$\{\tilde{h}_{(e1)A},$ $\tilde{h}_{(o1)A},$ $\tilde{h}_{(e0)},$
$\tilde{h}_{(e2)},$ $\tilde{h}_{(o2)}\}$ is guaranteed.
Because the eigenvalue of the derivative operator $\hat{\Delta}$ on
$S^{2}$ is $-l(l+1)$, the kernels of the operators $\hat{\Delta}$ and
$\hat{\Delta}+2$ are $l = 0$ and $l = 1$ modes, respectively.
Thus, the one-to-one correspondence between
$\{h_{Ap},$ $h_{pq}\}$ and
$\{\tilde{h}_{(e1)A},$ $\tilde{h}_{(o1)A},$ $\tilde{h}_{(e0)},$
$\tilde{h}_{(e2)},$ $\tilde{h}_{(o2)}\}$ is lost for $l = 0,1$ modes in
decomposition formulae
(\ref{eq:hAB-fourier})--(\ref{eq:hpq-fourier}) with
$S_{\delta}=Y_{lm}$.
To recover this one-to-one correspondence, in
Ref.~\cite{K.Nakamura-2021a}, we introduced the mode functions
$k_{(\hat{\Delta})}$ and $k_{(\hat{\Delta}+2)m}$ instead of $Y_{00}$
and $Y_{1m}$, respectively, and consider the scalar harmonic function
\begin{eqnarray}
  \label{eq:extended-harmonic-functions}
  S_{\delta} = \left\{
  \begin{array}{lcccl}
    Y_{lm} &\quad& \mbox{for} &\quad& l\geq 2; \\
    k_{(\hat{\Delta}+2)m} &\quad& \mbox{for} &\quad&  l=1; \\
    k_{(\hat{\Delta})} &\quad& \mbox{for} &\quad& l=0.
  \end{array}
                               \right.
\end{eqnarray}
As the explicit functions of $k_{(\hat{\Delta})}$ and
$k_{(\hat{\Delta}+2)m}$, we employ
\begin{eqnarray}
  \label{eq:l=0-general-mode-func-specific}
  &&
     k_{(\hat{\Delta})}
     =
     1
     +
     \delta \ln\left(\frac{1-z}{1+z}\right)^{1/2}
     ,
     \quad \delta\in\RF
     ,
  \\
  \label{eq:l=1-m=0-general-mode-func-specific}
  &&
     k_{(\hat{\Delta}+2)m=0}
     =
     z \left\{
     1
     +
     \delta
     \left(\frac{1}{2}\ln\frac{1+z}{1-z}-\frac{1}{z}\right)
     \right\}
     ,
  \\
  \label{eq:l=1-m=pm1-general-mode-func-specific}
  &&
     k_{(\hat{\Delta}+2)m=\pm 1}
     =
     (1-z^{2})^{1/2}
     \nonumber\\
  && \quad\quad\quad\quad\quad\quad
     \times
     \left\{
     1
     +
     \delta
     \left(\frac{1}{2}\ln\frac{1+z}{1-z}+\frac{z}{1-z^{2}}\right)
     \right\} e^{\pm i \phi}
     ,
\end{eqnarray}
where $z = \cos\theta$.
This choice guarantees the linear-independence of the set of the
harmonic functions
\begin{eqnarray}
  \label{eq:set-of-harmonic-functions}
  &&
     \left\{S_{\delta}, \hat{D}_{p}S_{\delta},
     \epsilon_{pq}\hat{D}^{q}S_{\delta},
     \displaystyle \frac{1}{2} \gamma_{pq}S_{\delta},
     \right.
     \nonumber\\
  && \quad\quad
     \left.
     \left(\hat{D}_{p}\hat{D}_{q}-\frac{1}{2}\gamma_{pq}\hat{D}^{r}\hat{D}_{r}\right)S_{\delta},
     2\epsilon_{r(p}\hat{D}_{q)}\hat{D}^{r}S_{\delta}\right\}
\end{eqnarray}
including $l=0,1$ modes if $\delta\neq 0$, but is singular if
$\delta\neq 0$.
When $\delta = 0$, we have $k_{(\hat{\Delta})}\propto Y_{00}$ and
$\hat{k}_{(\hat{\Delta}+2)m}\propto Y_{1m}$.


Using the above harmonics functions $S_{\delta}$ in
Eq.~(\ref{eq:extended-harmonic-functions}),
in Ref.~\cite{K.Nakamura-2021a}, we proposed the following strategy:
\begin{proposal}
  \label{proposal:harmonic-extension}
  We decompose the metric perturbations $h_{ab}$ on the background
  spacetime with the metric
  (\ref{eq:background-metric-2+2})--(\ref{eq:background-metric-2+2-gamma-comp-Schwarzschild}),
  through Eqs.~(\ref{eq:hAB-fourier})--(\ref{eq:hpq-fourier}) with the
  harmonic functions $S_{\delta}$ given by
  Eq.~(\ref{eq:extended-harmonic-functions}).
  Then, Eqs.~(\ref{eq:hAB-fourier})--(\ref{eq:hpq-fourier}) become
  invertible with the inclusion of $l=0,1$ modes.
  After deriving the field equations such as linearized Einstein
  equations using the harmonic function $S_{\delta}$, we choose
  $\delta=0$ when we solve these field equations as the regularity of
  the solutions.
\end{proposal}
Through this strategy, we can construct gauge-invariant variables and
evaluate field equations through the mode-by-mode analyses without
special treatments for $l = 0,1$ modes.


Once we accept Proposal~\ref{proposal:harmonic-extension}, we reach to
the following statement~\cite{K.Nakamura-2021a}:
\begin{theorem}
  \label{theorem:decomposition-theorem-Schwarzschild}
  If the gauge-transformation rule for a tensor field $h_{ab}$ is
  given by ${}_{{\cal Y}}\!h_{ab}$ $-$ ${}_{{\cal X}}\!h_{ab}$ $=$
  ${\pounds}_{\xi_{(1)}}g_{ab}$. Here, $g_{ab}$ is the background
  metric with the spherical symmetry.
  Then, there exist a tensor field ${\cal F}_{ab}$ and a vector
  field $Y^{a}$ such that $h_{ab}$ is decomposed as $h_{ab}$ $=:$
  ${\cal F}_{ab}$ $+$ ${\pounds}_{Y}g_{ab}$, where ${\cal F}_{ab}$ and
  $Y^{a}$ are transformed as ${}_{{\cal Y}}\!{\cal F}_{ab}$ $-$
  ${}_{{\cal X}}\!{\cal F}_{ab}$ $=$ $0$, ${}_{{\cal Y}}\!Y^{a}$
  $-$ ${}_{{\cal X}}\!Y^{a}$ $=$ $\xi^{a}_{(1)}$ under the gauge
  transformation.
\end{theorem}
Owing to Theorem~\ref{theorem:decomposition-theorem-Schwarzschild},
the above general arguments in our gauge-invariant perturbation theory
are applicable to perturbations on the Schwarzschild background
spacetime including $l=0,1$ mode perturbations.
Furthermore, we derived the $l=0,1$ solution to the linearized
Einstein equation in the gauge-invariant
manner~\cite{K.Nakamura-2021a}.


As shown in Eq.~(\ref{eq:nth-order-Einstein-eq}), the linearized
Einstein equation ${}^{(1)}\!G_{a}^{\;\;b}=8\pi
{}^{(1)}\!T_{a}^{\;\;b}$  for the linear metric perturbation
$h_{ab}={\cal F}_{ab}+{\pounds}_{Y}g_{ab}$ with the vacuum background
Einstein equation $G_{a}^{\;\;b}=8\pi T_{a}^{\;\;b}=0$ is given by
\begin{eqnarray}
  \label{eq:linear-Einstein-eq-gauge-inv}
  {}^{(1)}\!{\cal G}_{a}^{\;\;b}\left[{\cal F}\right]=8\pi {}^{(1)}\!{\cal T}_{a}^{\;\;b},
\end{eqnarray}
Since we consider the vacuum background spacetime $T_{ab}=0$, the
linear-order perturbation of the continuity equation of the
linear perturbation of the energy-momentum tensor is given by
\begin{eqnarray}
  \nabla^{a}{}^{(1)}\!{\cal T}_{a}^{\;\;b} = 0.
  \label{eq:divergence-barTab-linear-vac-back-u}
\end{eqnarray}
We decompose the components of the linear perturbation of
${}^{(1)}\!{\cal T}_{ac}$ as
\begin{eqnarray}
  &&
     \!\!\!\!\!\!\!\!\!\!\!\!\!\!\!\!\!
     {}^{(1)}\!{\cal T}_{ac}
     =
     \sum_{l,m}
     \tilde{T}_{AC}
     S_{\delta}
     (dx^{A})_{a} (dx^{C})_{c}
     \nonumber\\
  && \quad\quad
     +
      2
      r
      \sum_{l,m} \left\{
      \tilde{T}_{(e1)A} \hat{D}_{p}S_{\delta}
      +
      \tilde{T}_{(o1)A} \epsilon_{pq} \hat{D}^{q}S_{\delta}
      \right\}
      (dx^{A})_{(a} (dx^{p})_{c)}
      \nonumber\\
  && \quad\quad
     +
     r^{2}
     \sum_{l,m} \left\{
     \tilde{T}_{(e0)} \frac{1}{2} \gamma_{pq} S_{\delta}
     +
     \tilde{T}_{(e2)} \left(
     \hat{D}_{p}\hat{D}_{q}
     -
     \frac{1}{2} \gamma_{pq} \hat{D}_{r}\hat{D}^{r}
     \right) S_{\delta}
     \right.
     \nonumber\\
  &&
     \quad\quad\quad\quad\quad\quad
     \left.
     +
     \tilde{T}_{(o2)} \epsilon_{s(p}\hat{D}_{q)}\hat{D}^{s}S_{\delta}
     \right\}
     (dx^{p})_{a} (dx^{q})_{c}
     .
     \label{eq:1st-pert-calTab-dd-decomp-2}
\end{eqnarray}
We also derive the continuity equations
(\ref{eq:divergence-barTab-linear-vac-back-u}) in terms of these mode
coefficients and use these equations when we solve the linearized
Einstein equation.


Furthermore, we derived the solutions to the Einstein equation for
$l=0,1$ mode imposing the regularity of the harmonics $S_{\delta}$
through $\delta=0$.
For this reason, we may choose $\tilde{T}_{(e2)}$ $=$
$\tilde{T}_{(o2)}$ $=$ $0$ for $l=0,1$ modes.
In addition, we may also choose $\tilde{T}_{(e1)A}=0$ and
$\tilde{T}_{(o1)A}=0$ for $l=0$ modes due to the same reason.
This choice and a component of
Eq.~(\ref{eq:divergence-barTab-linear-vac-back-u}) leads
$\tilde{T}_{(e0)}=0$ for $l=0$ mode.


Through the above premise, in Ref.~\cite{K.Nakamura-2021a}, we derived
the $l=0,1$-mode solutions to the linearized Einstein equations as
follows:


For $l=1$ $m=0$ odd-mode perturbations, we derived
\begin{eqnarray}
  \label{eq:l=1-odd-mode-propagating-sol-ver2}
  &&
     \!\!\!\!\!\!\!\!\!\!\!\!\!\!\!\!\!\!\!\!\!\!\!
     2 {}^{(1)}\!{\cal F}_{Ap}(dx^{A})_{(a}(dx^{p})_{b)}
     \nonumber\\
  &\!\!\!\!\!\!\!\!\!\!\!=&\!\!\!\!\!\!\!\!\!\!
      \left(
      6M
      r^{2} \int dr
      \frac{1}{r^{4}} a_{1}(t,r)
      \right) \sin^{2}\theta (dt)_{(a}(d\phi)_{b)}
      +
      {\pounds}_{V_{(1,o1)}}g_{ab}
      ,
\end{eqnarray}
where the generator $V_{(1,o1)}^{a}$ of the term
${\pounds}_{V_{(1,o1)}}g_{ab}$ in
Eq.~(\ref{eq:l=1-odd-mode-propagating-sol-ver2}) is
\begin{eqnarray}
  \label{eq:l=1-odd-mode-propagating-sol-ver2-Va-def}
  V_{(1,o1)a}
  =
  \left(\beta_{1}(t) + W_{(1,o)}(t,r)\right) r^{2} \sin^{2}\theta (d\phi)_{a}
  .
\end{eqnarray}
Here, $\beta_{1}(t)$ is an arbitrary function of $t$.
The function $a_{1}(t,r)$ is given by as the solutions to the
linear-order Einstein equation (\ref{eq:linear-Einstein-eq-gauge-inv})
as follows:
\begin{eqnarray}
  a_{1}(t,r)
  &=&
  - \frac{16 \pi}{3M} r^{3} f \int dt \tilde{T}_{(o1)r} + a_{10}
      \nonumber\\
  &=&
  - \frac{16 \pi}{3M} \int dr r^{3} \frac{1}{f} \tilde{T}_{(o1)t} + a_{10}
  ,
  \label{eq:a1tr-sol}
\end{eqnarray}
where $a_{10}$ is the constant of integration which corresponds to the
Kerr parameter perturbation.
Furthermore $rf \partial_{r}W_{(1,o)}$ of the variable $W_{(1,o)}$ in
Eq.~(\ref{eq:l=1-odd-mode-propagating-sol-ver2-Va-def}) is determined
the evolution equation
\begin{eqnarray}
  &&
     \partial_{t}^{2}(r f \partial_{r}W_{(1,o)})
     -  f \partial_{r}( f \partial_{r}(r f \partial_{r}W_{(1,o)})
     \nonumber\\
  && \quad\quad
     + \frac{1}{r^{2}} f \left[ 3f-1 \right] (r f \partial_{r}W_{(1,o)})
     =
     16 \pi f^{2} \tilde{T}_{(o1)r}
     .
     \label{eq:odd-master-equation-Regge-Wheeler-l=1}
\end{eqnarray}


For the $l=0$ even-mode perturbation, we should have
\begin{eqnarray}
  &&
     {}^{(1)}\!{\cal F}_{ab}
     =
     \frac{2}{r}
     \left(
     M_{1}
     + 4 \pi \int dr \left[\frac{r^{2}}{f} \tilde{T}_{tt}\right]
     \right)
     \nonumber\\
  && \quad\quad\quad\quad\quad
     \times
     \left(
     (dt)_{a}(dt)_{b}
     +
     \frac{1}{f^{2}}
     (dr)_{a}(dr)_{b}
     \right)
     \nonumber\\
  && \quad\quad\quad\quad
     +
     2 \left[
     4 \pi r \int dt \left(
     \frac{1}{f} \tilde{T}_{tt}
     + f \tilde{T}_{rr}
     \right)
     \right]
     (dt)_{(a}(dr)_{b)}
     \nonumber\\
  && \quad\quad\quad\quad
     +
     {\pounds}_{V_{(1,e0)}}g_{ab}
     ,
     \label{eq:l=0-final-sols}
\end{eqnarray}
where $M_{1}$ is the linear-order Schwarzschild mass parameter
perturbation, $\gamma_{1}(r)$ is an arbitrary function of $r$.
Here, the generator $V_{(1,e0)a}$ of the term
${\pounds}_{V_{(1,e0)}}g_{ab}$ in Eq.~(\ref{eq:l=0-final-sols}) is
given by
\begin{eqnarray}
  \label{eq:Va-second-choice-non-vac-sum}
  &&
     V_{(1,e0)a}
     :=
     \left(
     \frac{1}{4} f \Upsilon_{1} + \frac{1}{4} r f \partial_{r}\Upsilon_{1}
     + \gamma_{1}(r)
     \right)
     (dt)_{a}
     \nonumber\\
  && \quad\quad\quad\quad\quad
     +
     \frac{1}{4f} r \partial_{t}\Upsilon_{1}
     (dr)_{a}
     ,
\end{eqnarray}
In the generator (\ref{eq:Va-second-choice-non-vac-sum}),
${}^{(1)}\!\tilde{F}:=\partial_{t}\Upsilon_{1}$ satisfies the
following equation:
\begin{eqnarray}
  &&
     -  \frac{1}{f} \partial_{t}^{2}\tilde{F}
     + \partial_{r}( f \partial_{r}\tilde{F} )
     + \frac{1}{r^{2}} 3(1-f) \tilde{F}
     \nonumber\\
  &=&
      - \frac{8}{r^{3}} m_{1}(t,r)
      + 16 \pi \left[
      -  \frac{1}{f} \tilde{T}_{tt}
      + f \tilde{T}_{rr}
      \right]
      ,
      \label{eq:even-mode-tildeF-eq-Phie-reduce}
\end{eqnarray}
where
\begin{eqnarray}
  m_{1}(t,r)
  &=&
      4 \pi \int dr \left[\frac{r^{2}}{f} \tilde{T}_{tt}\right]
      + M_{1}
      \nonumber\\
  &=&
      4 \pi \int dt \left[ r^{2} f \tilde{T}_{rt} \right]
      + M_{1}
      ,
      \quad
      M_{1}\in\RF
      .
      \label{eq:Ein-non-vac-m1-sol}
\end{eqnarray}


For the $l=1$ $m=0$ even-mode perturbation, we should have
\begin{eqnarray}
  &&
     {}^{(1)}\!{\cal F}_{ab}
     =
     - \frac{16\pi r^{2} f^{2}}{3(1-f)} \left[
     \frac{1+f}{2} \tilde{T}_{rr}
     + r f \partial_{r}\tilde{T}_{rr}
     -  \tilde{T}_{(e0)}
     \right.
     \nonumber\\
  && \quad\quad\quad\quad\quad\quad\quad\quad\quad\quad
     \left.
     -  4 \tilde{T}_{(e1)r}
     \right] \cos\theta (dt)_{a}(dt)_{b}
     \nonumber\\
  && \quad\quad\quad\quad
      +
      16 \pi r^{2} \left\{
      \tilde{T}_{tr}
      - \frac{2r}{3f(1-f)} \partial_{t}\tilde{T}_{tt}
      \right\}
     \nonumber\\
  && \quad\quad\quad\quad\quad\quad
     \times
     \cos\theta (dt)_{(a}(dr)_{b)}
     \nonumber\\
  && \quad\quad\quad\quad
     + \frac{8 \pi r^{2} (1-3f)}{f^{2}(1-f)} \left[
     \tilde{T}_{tt}
     -  \frac{2rf}{3(1-3f)} \partial_{r}\tilde{T}_{tt}
     \right]
     \nonumber\\
  && \quad\quad\quad\quad\quad\quad
     \times
     \cos\theta (dr)_{a}(dr)_{b}
     \nonumber\\
  && \quad\quad\quad\quad
     -  \frac{16 \pi r^{4}}{3(1-f)} \tilde{T}_{tt} \cos\theta \gamma_{ab}
     +
     {\pounds}_{V_{(1,e1)}}g_{ab}
      \label{eq:calFab-l=1-m=0-sol}
     ,
  \\
  &&
     V_{(1,e1)a}
     :=
     -  r \partial_{t}\Phi_{(e)} \cos\theta (dt)_{a}
     \nonumber\\
  && \quad\quad\quad\quad\quad
     + \left( \Phi_{(e)} - r \partial_{r}\Phi_{(e)} \right) \cos\theta (dr)_{a}
     \nonumber\\
  && \quad\quad\quad\quad\quad
     -  r \Phi_{(e)} \sin\theta (d\theta)_{a}
     ,
     \label{eq:generator-covariant-l=1-m=0-sum}
\end{eqnarray}
where $\Phi_{(e)}$ satisfies the following equation
\begin{eqnarray}
  &&
     \!\!\!\!\!\!\!\!\!\!\!\!\!\!
     -  \frac{1}{f} \partial_{t}^{2}\Phi_{(e)}
     + \partial_{r}\left[ f \partial_{r}\Phi_{(e)} \right]
     -
     \frac{1-f}{r^{2}} \Phi_{(e)}
     =
     16 \pi \frac{r}{3(1-f)} S_{(\Phi_{(e)})}
     ,
  \nonumber\\
  &&
     \!\!\!\!\!\!\!\!\!\!\!\!\!\!
     S_{(\Phi_{(e)})}
     :=
     \frac{3(1-3f)}{4f} \tilde{T}_{tt}
     -  \frac{1}{2} r \partial_{r}\tilde{T}_{tt}
     + \frac{1+f}{4} f \tilde{T}_{rr}
     + \frac{1}{2} f^{2} r \partial_{r}\tilde{T}_{rr}
     \nonumber\\
  && \quad\quad\quad
     -  \frac{f}{2} \tilde{T}_{(e0)}
     -  2 f \tilde{T}_{(e1)r}
     .
     \label{eq:SPhie-def-explicit-l=1}
\end{eqnarray}


\section{Extension to the higher-order\\ perturbations}
\label{sec:higher-order_extension}


As reviewed in Sec.~\ref{sec:review-of-perturbation-theroy}, the
$n$-th order perturbation of the Einstein equation is given in the
gauge-invariant form.
We may write this $n$-th order Einstein equation
(\ref{eq:nth-order-Einstein-eq}) as follows:
\begin{eqnarray}
  &&
     {}^{(1)}\!{\cal G}_{a}^{\;\;b}\left[{}^{(n)}\!{\cal F}\right]
     =
     -
     {}^{({\rm NL})}\!{\cal G}_{a}^{\;\;b}\left[
     \left\{ \left. {}^{(i)}\!{\cal F}_{cd}\right| i<n \right\}
     \right]
     +
     8 \pi {}^{(n)}\!{\cal T}_{a}^{\;\;b}
     \nonumber\\
  && \quad\quad\quad\quad\quad\quad
     =:
     8 \pi {}^{(n)}\!\TF_{a}^{\;\;b}
     .
     \label{eq:nth-einstein-gauge-inv-again}
\end{eqnarray}
Here, the left-hand side in
Eq.~(\ref{eq:nth-einstein-gauge-inv-again}) is the linear term of
${}^{(n)}\!{\cal F}_{ab}$ and the first term in the right-hand side
is the non-linear term consists of the lower-order metric
perturbation ${}^{(i)}\!{\cal F}_{ab}$ with $i<n$.
The right-hand side $8 \pi {}^{(n)}\!\TF_{a}^{\;\;b}$ of
Eq.~(\ref{eq:nth-einstein-gauge-inv-again}) is regarded an
effective energy-momentum tensor for the $n$-th order metric
perturbation ${}^{(n)}\!{\cal F}_{ab}$.


The vacuum background condition $G_{a}^{\;\;b}=0$ implies the
mathematical identity (\ref{eq:linear-perturbation-of-div-Gab-vacuum}),
and Eq.~(\ref{eq:nth-einstein-gauge-inv-again}) implies
\begin{eqnarray}
  \label{eq:nth-order-divergenceTFab}
  \nabla^{a}{}^{(n)}\!\TF_{a}^{\;\;b} = 0.
\end{eqnarray}
This equation gives consistency relations which should be confirmed in
concrete physical situations.
The first term in the right-hand side in
Eq.~(\ref{eq:nth-einstein-gauge-inv-again}) does not contain
${}^{(n)}{\cal F}_{ab}$.
The $n$-th order perturbation ${}^{(n)}{\cal T}_{a}^{\;\;b}$ does
not contain ${}^{(n)}{\cal F}_{ab}$, neither, because our background
spacetime is vacuum.
Then, ${}^{(n)}\!\TF_{a}^{\;\;b}$ does not include
${}^{(n)}{\cal F}_{ab}$.
This situation is same as that we used when we solved the linear-order
Einstein equation (\ref{eq:linear-Einstein-eq-gauge-inv}) with the
linear perturbation (\ref{eq:divergence-barTab-linear-vac-back-u}) of
the continuity equation of the energy-momentum
in Ref.~\cite{K.Nakamura-2021a}.
Furthermore, we decompose the tensor ${}^{(n)}\!\TF_{ab}$ as follows:
\begin{eqnarray}
  &&
     \!\!\!\!\!\!\!\!\!\!\!\!\!\!\!\!\!
     {}^{(1)}\!\TF_{ab}
     =:
     \sum_{l,m}
     \tilde{\TF}_{AB}
     S_{\delta}
     (dx^{A})_{a} (dx^{B})_{b}
     \nonumber\\
  && \quad\quad
     +
     2
     r
     \sum_{l,m} \left\{
     \tilde{\TF}_{(e1)A} \hat{D}_{p}S_{\delta}
     +
     \tilde{\TF}_{(o1)A} \epsilon_{pq} \hat{D}^{q}S_{\delta}
     \right\}
     (dx^{A})_{(a} (dx^{p})_{b)}
     \nonumber\\
  && \quad\quad
     +
     r^{2}
     \sum_{l,m} \left\{
     \tilde{\TF}_{(e0)} \frac{1}{2} \gamma_{pq} S_{\delta}
     +
     \tilde{\TF}_{(e2)} \left(
     \hat{D}_{p}\hat{D}_{q}
     -
     \frac{1}{2} \gamma_{pq} \hat{D}_{r}\hat{D}^{r}
     \right) S_{\delta}
     \right.
     \nonumber\\
  &&
     \quad\quad\quad\quad\quad\quad
     \left.
     +
     \tilde{\TF}_{(o2)} \epsilon_{s(p}\hat{D}_{q)}\hat{D}^{s}S_{\delta}
     \right\}
     (dx^{p})_{a} (dx^{q})_{b}
     .
     \label{eq:1st-pert-TFab-dd-decomp}
\end{eqnarray}
Then, the replacements
\begin{eqnarray}
  &&
     \tilde{T}_{AB}\rightarrow \tilde{\TF}_{AB}, \quad
     \tilde{T}_{(e1)A}\rightarrow \tilde{\TF}_{(e1)A}, \quad
     \tilde{T}_{(o1)A}\rightarrow \tilde{\TF}_{(o1)A},
  \nonumber\\
  &&
     \tilde{T}_{(e0)}\rightarrow \tilde{\TF}_{(e0)}, \quad
     \tilde{T}_{(e2)}\rightarrow \tilde{\TF}_{(e2)}, \quad
     \tilde{T}_{(o2)}\rightarrow \tilde{\TF}_{(o2)}
     \label{eq:tildeTab-tildeTFab-repalcements}
\end{eqnarray}
in the solutions
(\ref{eq:l=1-odd-mode-propagating-sol-ver2})--(\ref{eq:SPhie-def-explicit-l=1})
yield the solutions to Eq.~(\ref{eq:nth-einstein-gauge-inv-again}).


Then, following the strategy as
Proposal~\ref{proposal:harmonic-extension} and the results derived
in Ref.~\cite{K.Nakamura-2021a}, the $l=0,1$-mode solutions to
Eq.~(\ref{eq:nth-einstein-gauge-inv-again}) are summarized as follows:


For $l=1$ $m=0$ odd-mode perturbations, we should have
\begin{eqnarray}
  \label{eq:l=1-odd-mode-propagating-sol-ver2-nth}
  &&
     \!\!\!\!\!\!\!\!\!\!\!\!\!\!\!\!\!\!\!\!\!\!\!
     2 {}^{(n)}\!{\cal F}_{Ap}(dx^{A})_{(a}(dx^{p})_{b)}
     \nonumber\\
  &\!\!\!\!\!\!\!\!\!\!\!=&\!\!\!\!\!\!\!\!\!\!
      \left(
      6M
      r^{2} \int dr
      \frac{1}{r^{4}} a_{n}(t,r)
      \right) \sin^{2}\theta (dt)_{(a}(d\phi)_{b)}
      +
      {\pounds}_{V_{(n,o1)}}g_{ab}
      ,
\end{eqnarray}
where the generator $V_{(n,o1)}^{a}$ of the term
${\pounds}_{V_{(n,o1)}}g_{ab}$ in
Eq.~(\ref{eq:l=1-odd-mode-propagating-sol-ver2-nth}) is
\begin{eqnarray}
  \label{eq:l=1-odd-mode-propagating-sol-ver2-Va-def-nth}
  V_{(n,o1)a}
  =
  \left(\beta_{n}(t) + W_{(n,o)}(t,r)\right) r^{2} \sin^{2}\theta (d\phi)_{a}
  .
\end{eqnarray}
Here, $\beta_{n}(t)$ is an arbitrary function of $t$.
The function $a_{n}(t,r)$ is given by as the solutions to the
nth-order Einstein equation (\ref{eq:nth-einstein-gauge-inv-again}) as
follows:
\begin{eqnarray}
  a_{n}(t,r)
  &=&
  - \frac{16 \pi}{3M} r^{3} f \int dt {}^{(n)}\!\tilde{\TF}_{(o1)r} + a_{n0}
      \nonumber\\
  &=&
  - \frac{16 \pi}{3M} \int dr r^{3} \frac{1}{f} {}^{(n)}\!\tilde{\TF}_{(o1)t} + a_{n0}
  ,
  \label{eq:a1tr-sol-nth}
\end{eqnarray}
where $a_{n0}$ is the constant of integration which corresponds to the
Kerr parameter perturbation.
Furthermore $rf \partial_{r}W_{(n,o)}$ of the variable $W_{(n,o)}$ in
Eq.~(\ref{eq:l=1-odd-mode-propagating-sol-ver2-Va-def-nth}) is determined
the evolution equation
\begin{eqnarray}
  &&
     \partial_{t}^{2}(r f \partial_{r}W_{(n,o)})
     -  f \partial_{r}( f \partial_{r}(r f \partial_{r}W_{(n,o)})
     \nonumber\\
  && \quad\quad
     + \frac{1}{r^{2}} f \left[ 3f-1 \right] (r f \partial_{r}W_{(n,o)})
     =
     16 \pi f^{2} {}^{(n)}\!\tilde{\TF}_{(o1)r}
     .
     \label{eq:odd-master-equation-Regge-Wheeler-l=1-nth}
\end{eqnarray}


For the $l=0$ even-mode perturbation, we should have
\begin{eqnarray}
  &&
     {}^{(n)}\!{\cal F}_{ab}
     =
     \frac{2}{r}
     \left(
     M_{n}
     + 4 \pi \int dr \left[\frac{r^{2}}{f} {}^{(n)}\!\tilde{\TF}_{tt}\right]
     \right)
     \nonumber\\
  && \quad\quad\quad\quad\quad
     \times
     \left(
     (dt)_{a}(dt)_{b}
     +
     \frac{1}{f^{2}}
     (dr)_{a}(dr)_{b}
     \right)
     \nonumber\\
  && \quad\quad\quad\quad
     +
     2 \left[
     4 \pi r \int dt \left(
     \frac{1}{f} {}^{(n)}\!\tilde{\TF}_{tt}
     + f {}^{(n)}\!\tilde{\TF}_{rr}
     \right)
     \right]
     (dt)_{(a}(dr)_{b)}
     \nonumber\\
  && \quad\quad\quad\quad
     +
     {\pounds}_{V_{(n,e0)}}g_{ab}
     ,
     \label{eq:l=0-final-sols-nth}
\end{eqnarray}
where $M_{n}$ is the $n$th-order Schwarzschild mass parameter
perturbation, $\gamma_{n}(r)$ is an arbitrary function of $r$.
Here, the generator $V_{(n,e0)a}$ of the term
${\pounds}_{V_{(n,e0)}}g_{ab}$ in Eq.~(\ref{eq:l=0-final-sols-nth}) is
given by
\begin{eqnarray}
  \label{eq:Va-second-choice-non-vac-sum-nth}
  &&
     V_{(n,e0)a}
     :=
     \left(
     \frac{1}{4} f \Upsilon_{n} + \frac{1}{4} r f \partial_{r}\Upsilon_{n}
     + \gamma_{n}(r)
     \right)
     (dt)_{a}
     \nonumber\\
  && \quad\quad\quad\quad\quad
     +
     \frac{1}{4f} r \partial_{t}\Upsilon_{n}
     (dr)_{a}
     ,
\end{eqnarray}
In the generator (\ref{eq:Va-second-choice-non-vac-sum-nth}),
${}^{(n)}\!\tilde{F}:=\partial_{t}\Upsilon_{n}$ satisfies the
following equation:
\begin{eqnarray}
  &&
     -  \frac{1}{f} \partial_{t}^{2}{}^{(n)}\!\tilde{F}
     + \partial_{r}( f \partial_{r}{}^{(n)}\!\tilde{F} )
     + \frac{1}{r^{2}} 3(1-f) {}^{(n)}\!\tilde{F}
     \nonumber\\
  &=&
      - \frac{8}{r^{3}} m_{n}(t,r)
      + 16 \pi \left[
      -  \frac{1}{f} {}^{(n)}\!\tilde{\TF}_{tt}
      + f {}^{(n)}\!\tilde{\TF}_{rr}
      \right]
      ,
      \label{eq:even-mode-tildeF-eq-Phie-reduce-nth}
\end{eqnarray}
where
\begin{eqnarray}
  m_{n}(t,r)
  &=&
      4 \pi \int dr \left[\frac{r^{2}}{f} {}^{(n)}\!\tilde{\TF}_{tt}\right]
      + M_{n}
      \nonumber\\
  &=&
      4 \pi \int dt \left[ r^{2} f {}^{(n)}\!\tilde{\TF}_{rt} \right]
      + M_{n}
      ,
      \quad
      M_{n}\in\RF
      .
      \label{eq:Ein-non-vac-m1-sol-nth}
\end{eqnarray}


For the $l=1$ $m=0$ even-mode perturbation, we should have
\begin{eqnarray}
  &&
     {}^{(n)}\!{\cal F}_{ab}
     =
     - \frac{16\pi r^{2} f^{2}}{3(1-f)} \left[
     \frac{1+f}{2} {}^{(n)}\!\tilde{\TF}_{rr}
     + r f \partial_{r}{}^{(n)}\!\tilde{\TF}_{rr}
     -  {}^{(n)}\!\tilde{\TF}_{(e0)}
     \right.
     \nonumber\\
  && \quad\quad\quad\quad\quad\quad\quad\quad\quad\quad
     \left.
     -  4 {}^{(n)}\!\tilde{\TF}_{(e1)r}
     \right] \cos\theta (dt)_{a}(dt)_{b}
     \nonumber\\
  && \quad\quad\quad\quad
      +
      16 \pi r^{2} \left\{
      {}^{(n)}\!\tilde{\TF}_{tr}
      - \frac{2r}{3f(1-f)} \partial_{t}{}^{(n)}\!\tilde{\TF}_{tt}
      \right\}
     \nonumber\\
  && \quad\quad\quad\quad\quad\quad
     \times
     \cos\theta (dt)_{(a}(dr)_{b)}
     \nonumber\\
  && \quad\quad\quad\quad
     + \frac{8 \pi r^{2} (1-3f)}{f^{2}(1-f)} \left[
     {}^{(n)}\!\tilde{\TF}_{tt}
     -  \frac{2rf}{3(1-3f)} \partial_{r}{}^{(n)}\!\tilde{\TF}_{tt}
     \right]
     \nonumber\\
  && \quad\quad\quad\quad\quad\quad
     \times
     \cos\theta (dr)_{a}(dr)_{b}
     \nonumber\\
  && \quad\quad\quad\quad
     -  \frac{16 \pi r^{4}}{3(1-f)} {}^{(n)}\!\tilde{\TF}_{tt} \cos\theta \gamma_{ab}
     +
     {\pounds}_{V_{(n,e1)}}g_{ab}
      \label{eq:calFab-l=1-m=0-sol.-nth}
     ,
  \\
  &&
     V_{(n,e1)a}
     :=
     -  r \partial_{t}\Phi_{(n,e)} \cos\theta (dt)_{a}
     \nonumber\\
  && \quad\quad\quad\quad\quad
     + \left( \Phi_{(n,e)} - r \partial_{r}\Phi_{(n,e)} \right) \cos\theta (dr)_{a}
     \nonumber\\
  && \quad\quad\quad\quad\quad
     -  r \Phi_{(n,e)} \sin\theta (d\theta)_{a}
     .
     \label{eq:generator-covariant-l=1-m=0-sum-2-nth}
\end{eqnarray}


These are the main assertion of this article.


\section{Summary}
\label{sec:Summary}


In summary, we extended the linear-order solution of the mass
perturbation ($l=0$ even mode), the angular-momentum perturbation
($l=1$ odd mode), and the dipole perturbation ($l=1$ even mode)
to the any-order formal solutions.
Our logic starts from the complete proof of
Conjecture~\ref{conjecture:decomposition-conjecture} for perturbations
on the Schwarzschild background spacetime.
The remaining problem in
Conjecture~\ref{conjecture:decomposition-conjecture} was in the
treatment of $l=0,1$ modes of the perturbations on the Schwarzschild
background spacetime.
To resolve this problem, in Ref.~\cite{K.Nakamura-2021a}, we
introduced the harmonic functions $S_{\delta}$ defined by
Eq.~(\ref{eq:extended-harmonic-functions}) instead of the conventional
harmonic function $Y_{lm}$ and proposed
Proposal~\ref{proposal:harmonic-extension} as a strategy of a
gauge-invariant treatment of the $l=0,1$ perturbations on the
Schwarzschild background spacetime.
Once we accept this proposal, we reach to
Theorem~\ref{theorem:decomposition-theorem-Schwarzschild} and we can
apply our general arguments of higher-order perturbation theory
developed in
Refs.~\cite{K.Nakamura-2003,K.Nakamura-2005,K.Nakamura-2011,K.Nakamura-2014}
to perturbations on the Schwarzschild background spacetime.


In Ref.~\cite{K.Nakamura-2021a}, we derived the $l=0,1$ solutions
(\ref{eq:l=1-odd-mode-propagating-sol-ver2})--(\ref{eq:SPhie-def-explicit-l=1})
to the linearized Einstein equations following
Proposal~\ref{proposal:harmonic-extension}.
The premise and equations for any-order perturbations are same as
those for the linear perturbations.
Then, we reached to the formal solutions
(\ref{eq:l=1-odd-mode-propagating-sol-ver2-nth})--(\ref{eq:generator-covariant-l=1-m=0-sum-2-nth})
for the any-order non-linear perturbation by the replacements
(\ref{eq:tildeTab-tildeTFab-repalcements}).


Of course, the solutions derived here is just formal one and we have
to evaluate the non-linear terms in the effective energy-momentum
tensor ${}^{(n)}\!\TF_{a}^{\;\;b}$, i.e.,
${}^{({\rm NL})}\!{\cal G}_{a}^{\;\;b}[\{ {}^{(i)}\!{\cal F}_{cd}| i<n\}]$ 
and ${}^{(n)}\!{\cal T}_{a}^{\;\;b}$.
This evaluation will depend on the situations which we want to
clarify.
In addition to the perturbations on the Schwarzschild background
spacetime, the strategy in Proposal~\ref{proposal:harmonic-extension}
is a clue of the generalization of applications of our general
framework on the gauge-invariant higher-order perturbations to other
physical situations such as higher-order gauge-invariant cosmological
perturbations~\cite{K.Nakamura-2006}.
We leave further evaluations of our formal solutions
(\ref{eq:l=1-odd-mode-propagating-sol-ver2-nth})--(\ref{eq:generator-covariant-l=1-m=0-sum-2-nth})
in specific physical situations and the applications to the other
perturbation theories with different background spacetimes as future
works.



\end{document}